\documentclass[10pt]{article}
\usepackage[dvips]{graphicx}

\usepackage{epsfig}
\usepackage{graphicx}
\usepackage{indentfirst}
\usepackage{amsmath}
\usepackage{amsfonts}
\usepackage{amssymb}

\begin{document}

\begin{center}
{\Large\bf   Quantum effects  from particle production on background evolution and  Cardy-Verlinde formula in f(R) gravity\\}
\medskip

 M. J. S. Houndjo\footnote{e-mail:
sthoundjo@yahoo.fr, 
 \,Instituto de F\'{i}sica, Universidade Federal da Bahia, 40210-340, Salvador, BA, Brazil},\,\,\,A. V. Monwanou\footnote{e-mail: movins2008@yahoo.fr,\,Institut de Math\'{e}matiques et de Sciences Physiques (IMSP), 01 BP 613 Porto-Novo, B\'{e}nin}\,\,and Jean B. Chabi Orou\footnote{e-mail: jean.chabi@imsp-uac.org,\,Institut de Math\'{e}matiques et de Sciences Physiques (IMSP), 01 BP 613 Porto-Novo, B\'{e}nin }
\medskip\\
   
\end{center}

\begin{abstract}
We investigate particle production in an expanding universe under the assumption that the Lagrangian contains the Einstein term $R$ plus a modified gravity term of the form $R^\alpha$, where $\alpha$  is a constant. Dark fluid is considered as the main content of the universe and the big rip singularity appears. Quantum effects due to particle creation is analysed near the singularity and we find that for $\alpha\in ]\frac{1}{2}, 1[$, quantum effects are dominant and the big rip may be avoided whereas for $\alpha\in J$ the dark fluid is dominant  and the singularity remains. The Cardy-Verlinde formula  is also introduced and its equivalence with the total entropy of the universe is checked. It is found that this can always occur in Einstein gravity while in f(R) gravity, it holds  only for $\alpha=\frac{n+1}{2n+6}$, $n$ being the space dimension, corresponding to the situation in which the big rip cannot be avoided.
\end{abstract}

Pacs numbers: 04.50.Kd, 98.80.Cq, 02.40.Xx, 05. 70.-a

\section{Introduction}
General relativity (GR) is widely accepted as a fundamental theory to describe the geometric properties of spacetime \cite{alejandro1}. In an homogeneous and isotropic spacetime, the so-called Friedmann-Robertson-Walker (FRW) model, the Einstein field equations give rise to the Friedmann equations that describe the evolution of the universe. It describes the universe from around one second after the big bang to the present matter dominated era.  This standard model's success is in part due to several of its predictions having been verified by observation. For example, the abundances Helium with respect to other light elements observed in the universe agrees well with the predictions of this model. The period of recombination is strongly supported by the CMB which is arguably the strongest evidence supporting the standard model.  A good amount of observational data indicate quite  clearly that the present universe is in an accelerated expanding phase \cite{brasil5, brasil6, psde1}. The universe may enter in a so-called phantom era  with an effective equation of state parameter $\omega$ less than $-1$. The simplest way to explain this phantom dark energy era is based on the introduction of a scalar field with negative kinetic energy \cite{psde21, psde22, psde23, psde24,psde25, psde26, psde27, psde28, psde29, psde210, psde211, psde212, psde213, psde214}. The main property of such a phantom field in the accelerating  FRW universe is the appearance of finite-time future singularity \cite{iver12, iver14} (see \cite{psde3, articlesudden, nojiriclass} for the classification of singularities). In turn, this can lead to bizarre consequences such as negative entropy of the universe \cite{iver15}. \par

However, GR is not the only relativistic theory of gravity. In the last decades several generalizations of Einstein  field equations has been proposed \cite{alejandro11, alejandro14}. Within these extended theories of gravity nowadays, a subclass known as $f(R)$ theory are an alternative for classical problems, as the accelerated expansion of the universe, instead of the dark energy and quintessence models \cite{alejandro15, alejandro21, mgarc1}.  Then,  an effective phantom phase can be realized without a scalar phantom. $f(R)$ theories of gravity are basically extensions of the Einstein-Hilbert action with an arbitrary function of Ricci scalar $R$ \cite{alejandro22, alejandro25, antonio77, antonio512, antonio102, antonio106}.  There are three formalisms in deriving field equations from the action in f(R) gravity. The first is the standard metric formalism in which the field equations are derived by the variation of the action with respect to the metric tensor $g_{\mu\nu}$. In this formalism the affine connection $\Gamma^{\lambda}_{\mu\nu}$ depends on $g_{\mu\nu}$. The second is the Palatini formalism \cite{antonio481} in which the metric and the connection are treated as independent variables when the action is varied. The third is the metric-affine formalism is which the gravitational action is a general function of the scalar curvature while the matter action is allowed to depend also on the connection \cite{alejandro25}. Note that these approaches give different field equations in f(R) gravity, while for the GR action they are identical with each other. The methodology leads to a boundary contribution which is usually dropped out setting null fluxes through Gauss-Stokes theorem \cite{alejandro7, alejandro8}.  In this paper we will use the metric formalism for obtaining the field equations.\par

As we mentioned above, in Einstein gravity, when the dark energy is introduced for explaining the late acceleration of the universe, finite time singularities can appear. Note that even in f(R) gravity where we do not need to introduce the dark energy for explaining the late acceleration,  finite time singularities can also appear in a background driven by a dark fluid. The question to be asked is: does particle creation can avoid these singularities or modify their nature? Particle production phenomenon in f(R) is then analysed in this paper and its impact as quantum effects is checked at singularity time. 
\par
Particle production from the vacuum by the gravitational field using quantum field theory in an expanding universe has been introduced firstly by Parker in the 1960s \cite{parker}. One of the interesting results  of this work is that in a radiative dominated expanding universe there is no massless particle creation due to the conformal invariance of the metric. 
 Latter, quantum process of  particle production has been studied by several authors, done in the course of the cosmological expansion \cite{ademir13, ademir14, ademir15, ademir16, ademir17, ademir18, ademir19, ademir20, ademir21, ademir22, ademir23}.  Recently,  various investigations in the aim of dealing with singularities have been done. Batista and collaborators \cite{article1}  studied the effects of particle creation when a massless scalar field is minimally coupled with the Einstein gravity. They found that the energy density of created particles never dominates the phantom energy density. In the same way, quantum effects near the big rip are studied in \cite{article3} where they used the $n$-wave regularization for calculating the energy density of particle creation and found that, in this case, it tends to infinity when the big rip is approached and becomes the dominant component of the universe. This means that big rip can be avoided. Pavlov \cite{pavlov}  computed both the number density of created particles and the stress-energy tensor for a conformally coupled massive field for the case in which $\omega=-5/3$. He found that quantum effects are not important if the field mass is much smaller than the Planck mass and the time left to the big rip is greater than the Planck time. Bates and Anderson \cite{anderson} used a background field approach in which the energy densities of the quantized fields are computed in the background spacetime which contains the big rip singularity. They found that for fields in realistic states for which the energy density of the quantized fields is small compared to that of the phantom energy density at early times, and for spacetimes with realistic values of $\omega$, there is no evidence that quantum effects become large enough to significantly affect the expansion of the spacetime until the spacetime curvature is of the order of the Planck scale or larger, at which point the semi classical approximation breaks down. \par
The calculations of particle production is usually done comparing the particle number at asymptotically early times, or with respect to vacuum state defined in two different frames. However, treating of quantizaton in curved space the main problem concerns the field theory interpretation in terms of particles. Note that in curved spacetime there is no Poincar\'{e} group symmetry and  the notion of vacuum becomes ambiguous. The problem may be solved using the diagonalization method of instantaneous Hamiltonian by a Bogoliubov transformation which leads to a finite results for the created particles number \cite{ademir22}.  Particles are created because the modes with positive and negative frequency
of the field become mixed during the universe expansion. Particle production is directly connected with the curvature of the universe and then when the field equation is put in the form of harmonic oscillator equation without friction part, its effect appears in the effective mass. Hence, in the radiative dominated universe with massless scalar field, either of zero or non-conformal coupling, the effective mass vanishes and the modes are not mixed in the course of the expansion due to conformal invariant of the metric and as consequence there is no creation of massless particles. Note also that in the static universe where the scale factor is constant and the energy density of ordinary matter is null, even the field is massive and minimally or conformally coupled with Einstein gravity, the effective mass vanishes and once again there is no particle production. However, even the density of ordinary matter is null and the modified gravity is taken into account, this could not hold. Under some assumptions the scale in such a condition could not be constant and as consequence, particle could be created. \par

In this paper we use the ansatz that the function $f(R)=R+\lambda R^\alpha $ and analyse the phenomenon of particle creation. In a first step we verify a known  result that the particle production is the same  as in Einstein gravity and this is seen through the scale factor behaviour. We consider the main content of the universe as the dark fluid and analyse particle production phenomenon. Note that with this dark fluid driven the background,  the universe may present the big rip. Quantum effects from particle creation is then checked near this singularity, comparing the classical energy density of the dark fluid with the renormalised energy density from the stress tensor corresponding to a massless minimally couple scalar field. We find that depending on the values of the parameter $n$, the big rip can be avoided. Then, we find that for $\alpha \in ]\frac{1}{2},1[$ the big rip can be removed whereas for $\alpha\in J$ it remains.

 In the second step an attention is devoted to the Cardy-Verlinde formula based on inhomogeneous equation of state. Verlinde  \cite{verlinde} made an interesting proposal that Cardy formula \cite{cardy} in two-dimensional conformal field theory can be generalized to arbitrary spacetime dimensions. Verlinde further proposed that a closed universe has subextensive (Casimir) contribution to its energy and entropy with the Casimir energy conjectured to be bounded from above by the Bekenstein-Hawking energy and as consequence, one obtains a very deep relation between gravity and thermodynamics \cite{youm}. Within the context of the radiation dominated universe, such bound on the Casimir energy is shown to lead to the Hubble and the Bekenstein entropy bounds respectively for the strongly and the weakly self-gravitating universes. The generalized entropy formula, called the CV formula, is further shown to coincide with the total entropy of the universe coming from the Friedmann equations. These results were later generalized \cite{group1,group2,group3,group4,group5,group6,group7,group8}. Our goal here is to analyse the equivalence between the CV formula and the total entropy coming from  the generalized Friedmann equations in f(R)  theory. Hence, we find that for the inhomogeneous EoS, the CV formula can always be recovered from the generalized entropy of the universe in Einstein gravity but only for $\alpha=\frac{n+1}{2n+6}$ in f(R) gravity. Moreover we check the behaviour of the CV formula terms near the singularity and find particularly that the Casimir energy decreases as the universe  expands  and vanishes near the singularity.

\par
The paper is organized as follows. In the second section a breve notion on particle production in expanding universe is addressed with the use of the diagonalization of instantaneous Hamiltonian by Bogoliubov transformation. In the third section we present the view of modified gravity about particle production and also quantum effects near the big rip are analysed. In the fourth section the equivalence between the total entropy of the universe and the CV formula is checked and an analysis of the quantum effects on CV formula terms near the singularity is also investigated.
The conclusions and perspectives are presented in the final section.

\section{Particle production in expanding background}

Let $\phi$ denotes a scalar field of mass m, and $a(\eta)$  the scale factor for  spatially flat homogeneous and isotropic FRW spacetime with conformal line element
\begin{eqnarray}\label{vincent1}
ds^2=a^2(\eta)\left( d\eta^2-dx^2-dy^2-dz^2\right)\,\,.
\end{eqnarray}
The Lagrangian density for this massive scalar field minimally coupled with  the gravitational field in a conformal spacetime reads
\begin{eqnarray}\label{vincent2}
\mathcal{L}=\frac{1}{2}a^2\eta^{\mu\nu}(\partial_\mu\phi)(\partial_\nu\phi)-\frac{1}{2}a^4m^2\phi^2\quad,
\end{eqnarray}
where $\eta_{\mu\nu}$ is Minkowski metric. The corresponding  field equation reads
\begin{eqnarray}\label{vincent3}
\frac{1}{a^2}\eta^{\mu\nu}\partial_{\mu}(a^2\partial_\nu)+a^2m^2\phi=0\quad.
\end{eqnarray}
We can decompose the real scalar field into the modes as 
\begin{eqnarray}\label{vincent4}
\phi(\eta,\vec{x})=\frac{1}{a}\int \frac{d^3k}{(2\pi)^{3/2}} e^{-i\vec{k}\vec{x}}\left[\chi_k(\eta)a_{\vec k}+\chi^{\ast}_{k}(\eta)a^{\dagger}_{-\vec k}\right]\quad.
\end{eqnarray}
The mode functions now satisfy the equation
\begin{eqnarray}\label{vincent5}
\chi^{\prime\prime}_k+\left(k^2+m^2a^2-\frac{a^{\prime\prime}}{a}\right)\chi_k=0\quad,
\end{eqnarray}
where the prime denotes the derivative with respect to the conformal time $\eta$.
The modes $\chi_k$ satisfy to the Wronskian relation
\begin{eqnarray}\label{vincent6}
\chi_k(\eta)\chi^{\ast\,\prime}_{k}(\eta) - \chi^{\ast}(\eta)\chi^{\prime}_{k}(\eta)=i\quad.
\end{eqnarray}
The system is quantized in a standard fashion by treating the field $\chi$ as an operator, imposing the equal-time commutation relations
\begin{eqnarray}\label{vincent7}
\left[\chi(\eta,\vec{x}),\pi(\eta,\vec{x}^{\,\prime})\right]= i\delta^3(\vec{x}-\vec{x}^{\,\prime})\,\,,
\end{eqnarray}
where $\pi=d\chi/d\eta\equiv \chi^{\,\prime}$ is the canonical momentum. Then, the operators $a_{\vec{k}}$ and $a_{\vec{k}}^{\dagger}$ satisfy the usual commutation relations
\begin{eqnarray}\label{vincent8}
[a_{\vec{k}},a^{\dagger}_{\vec{k}^{\prime}}]=\delta^3(\vec{k}-\vec{k}^{\,\prime})\,\quad,\quad[a_{\vec{k}},a_{\vec{k}^{\prime}}]=[a^{\dagger}_{\vec{k}},a^{\dagger}_{\vec{k}^{\prime}}]=0 \,\,.
\end{eqnarray}
The vacuum state is then defined as the state for which
\begin{eqnarray}\label{vincent9}
a_{\vec{k}}\left|0\right\rangle=0\quad\forall\,\, k\quad.
\end{eqnarray}
 Other states are built up from this by acting on it various combinations of creation operators $a^{\dagger}_{\vec{k}}$.\par
The Hamiltonian corresponding to  (\ref{vincent2}) is 
\begin{eqnarray}\label{vincent10}
H=\frac{1}{2}\int d^3x\left\{a^2\left[(\phi^{\prime})^2+(\nabla \phi)^2\right]+a^4m^2\phi^2\right\}\quad.
\end{eqnarray}
Making use of (\ref{vincent4}), the Hamiltonian can be re-written in terms of creation and annihilation operators and the mode functions as 
\begin{eqnarray}\label{vincent11}
H=\frac{1}{2}\int d^3k \left[ P_k(\eta)a_{\vec k}a_{-\vec k}+P^{\ast}_{k}(\eta)a^{\dagger}_{\vec k}a^{\dagger}_{-\vec k}+Q_{k}(\eta)\left(a_{\vec k}a^{\dagger}_{\vec k}+a^{\dagger}_{\vec k}a_{\vec k}\right)\right]\quad,
\end{eqnarray}
where $P_k(\eta)$ and $Q_{k}(\eta)$ are defined by 
\begin{eqnarray}
P_k&=&\left(-\frac{a^{\prime}}{a}\chi_k+\chi^{\prime}_k\right)^2+\omega_k^2\chi_k^2\quad, \label{vincent12}\\
Q_k&=&\left(-\frac{a^{\prime}}{a}\chi_k+\chi^{\prime}\right)\left(-\frac{a^\prime}{a}\chi^{\ast}_{k}+\chi^{\ast\,\prime}_{k}\right)+\omega^{2}_{k}\chi_k\chi^{\ast}_{k} \quad,\label{vincent13}
\end{eqnarray}
with $\omega_k^2=k^2+m^2a^2 $.\par
 In general, the field may be decomposed into many different complete set of modes and each of these  has its own vacuum state. Suppose we label one such set of modes as $\bar{\chi}(\eta)$. Then, in terms of these modes the field is
\begin{eqnarray}\label{vincent14}
\phi(\eta,\vec{x})=\frac{1}{a}\int \frac{d^3k}{(2\pi)^{3/2}} e^{-i\vec{k}\vec{x}}\left[\bar{\chi}_k(\eta)\bar{a}_{\vec k}+\bar{\chi}^{\ast}_{k}(\eta)\bar{a}^{\dagger}_{-\vec k}\right]\quad,
\end{eqnarray}
and the vacuum state is defined by $\bar{a}_{\vec{k}}\left|0\right\rangle=0$ for all $\vec{k}$. The Hamiltonian in this case is written as
\begin{eqnarray}\label{vincent15}
H=\frac{1}{2}\int d^3k \left[ \bar{P}_k(\eta)\bar{a}_{\vec k}\bar{a}_{-\vec k}+\bar{P}^{\ast}_{k}(\eta)\bar{a}^{\dagger}_{\vec k}\bar{a}^{\dagger}_{-\vec k}+\bar{Q}_{k}(\eta)\left(\bar{a}_{\vec k}\bar{a}^{\dagger}_{\vec k}+\bar{a}^{\dagger}_{\vec k}\bar{a}_{\vec k}\right)\right]\quad,
\end{eqnarray}
where $\bar{P}_k$ and $\bar{Q}_k$ are similar expressions as in  (\ref{vincent12}) and (\ref{vincent13}) respectively, replacing $\chi_k$ by $\bar{\chi}_k$. Because of completeness, the two sets of modes $\chi_{k}$ and $\bar{\chi}_{k}$ are related by the Bogolubov transformations that diagonalize the Hamiltonian i.e., $\bar{P}_k(\eta)=0$, satisfying the  relation
\begin{eqnarray}\label{vincent16}
\bar{\chi}_{k}(\eta)=\gamma_k\chi_{k}(\eta)+\beta_k\chi^{\ast}_{k}(\eta)\,\,,
\end{eqnarray}
with the  normalization condition $|\gamma_k(\eta)|^2-|\beta_k(\eta)|^2=1$, where $\gamma_k$ and $\beta_k$ are constants and called Bogolubov coefficients. One can compare the two vacuum states by noting that the number operator for the barred states is $\bar{N}\equiv \int d^3k\bar{a}^{\dagger}_{\vec{k}}\bar{a}_{\vec{k}}$. Taking its expectation value with respect to the unbarred vacuum, one finds
\begin{eqnarray}\label{vincent17}
\left\langle 0\right|\bar{N}\left|0\right\rangle = \int d^3k\left|\beta_k\right|^2\,\,.
\end{eqnarray}
Thus, the number of barred particles in the unbarred vacuum in the mode $\vec{k}$ is $\left|\beta_k\right|^2$. Similarly, the number of unbarred particles in the barred vacuum in the mode $\vec{k}$ is $\left|\beta_k\right|^2$. Since the diagonalization imposes $\bar{P}_k(\eta)=0$, using (\ref{vincent12}) with $\chi_k$ replaced by $\bar{\chi}_k$ , we get
\begin{eqnarray}\label{vincent18}
-\frac{a^{\prime}}{a}\bar{\chi}_k+\bar{\chi}^{\,\prime}_k=-i\omega_k\bar{\chi}_k\quad.
\end{eqnarray}
Substituting this result in the expression of $\bar{Q}_k(\eta)$, we get
\begin{eqnarray}\label{vincent19}
\bar{Q}_k(\eta) = 2\omega_{k}(\eta)|\bar{\chi}_k|^2\quad.
\end{eqnarray}
Using (\ref{vincent11}) and (\ref{vincent15}) we obtain 
\begin{eqnarray}\label{vincent20}
\left\langle 0\right|2\bar{a}^{\dagger}_{\vec k}(\eta)\bar{a}_{\vec k}(\eta)+1\left|0\right\rangle =\frac{Q_k(\eta)}{\bar{Q}_k(\eta)}\quad,
\end{eqnarray}
and consequently 
\begin{eqnarray}\label{vincent21}
|\beta_k(\eta)|^2= \frac{1}{2}\frac{Q_k(\eta)}{\bar{Q}_k(\eta)}-\frac{1}{2}\quad.
\end{eqnarray}
Using now (\ref{vincent13}) and (\ref{vincent19})  one can rewrite  (\ref{vincent21}) as 
\begin{eqnarray}\label{vincent22}
|\beta_k(\eta)|^2=\frac{1}{4}\frac{\left[\left(\frac{a^\prime}{a}\right)^2+\omega^2_k(\eta)\right]|\chi_k |^2-\frac{a^\prime}{a}\left(\chi_k\chi^{\ast\,\prime}_k+\chi^{\prime}_{k}\chi^{\ast}_{k}\right)+|\chi^{\prime}_{k}|^2}{\omega^2_k(\eta)|\bar{\chi}_k|^2}-\frac{1}{2}\quad.
\end{eqnarray}
There are two important point to be understood with respect to vacuum states in curved space. The first is that, in general,  it is uncertain what criteria should be used in choosing the vacuum state. The problem is that many of criteria used for Minkowski space as Lorentz invariance and positive frequency with respect to a timelike Killing vector no longer apply in curved space. The second one is that, if there is some "natural" choice of vacuum when the spacetime begins, it does not in general correspond to the natural choice of vacuum when the spacetime ends. That is, the "in" vacuum state and the "out" vacuum state are different. This leads to particle production as can be seen from (\ref{vincent22}).\par Now, to calculate $|\beta_k(\eta)|^2$  we have to find $\chi_k(\eta)$ from (\ref{vincent5}) and $\bar{\chi}_k(\eta)$ from (\ref{vincent18}), but this will be done explicitly in the next section where we will analyse the particle creation phenomenon when the scalar field is minimally coupled with the modified gravity.\par
With the metric (\ref{vincent1}) the Einstein equation leads to Friedmann ones as
\begin{eqnarray}
\frac{\left(a^{\prime}\right)^{2}}{a^{2}}=\frac{\kappa^2}{3}\rho \,a^{2} \label{vincent23}\quad ,\\
2\frac{a^{\prime\prime}}{a}-\frac{\left(a^{\prime}\right)^{2}}{a^{2}}=-\kappa^2 p\, a^{2} \label{vincent24}
\end{eqnarray}
where $\rho$ and $p$ are the energy density and the pressure respectively, and $\kappa^2=8\pi G$. We suppose that they are related by a barotropic equation of state such that
\begin{equation}
\label{vincent25}
p = \omega\rho \quad ,
\end{equation}
with $\omega$ the barotropic parameter. Combining Eqs. (\ref{vincent23}), (\ref{vincent24}) and (\ref{vincent25}), one obtains for the scale factor $a(\eta)\propto \eta^{\frac{2}{1+3\omega}}$ for the conformal time and $a(t)\propto t^{\frac{2}{3(1+\omega)}}$ for the cosmic time.\par
Our goal is to analyse particle creation phenomenon in f(R) gravity and check its impact as quantum effects on a possible appearance of singularities in the classical background when this is essentially driven by a dark fluid.

\section{Particle creation aspect in f(R) gravity}
In this case the curvature $R$ in Einstein Hilbert action is replaced  by  a function  f(R) and the total action reads
\begin{eqnarray}\label{vincent26}
S= \frac{1}{2\kappa^2}\int d^4x\sqrt{-g}\left[f(R)+\mathcal{L}(g_{\mu\nu},\phi)\right]\quad,
\end{eqnarray}
where $G$ is the gravitational constant, $g$ the determinant of the metric  $g_{\mu\nu}$ and $\mathcal{L}$ the matter Lagrangian which depends on the metric and the matter field $\phi$. The field equation can be derived varying the action (\ref{vincent26}) with respect to $g_{\mu\nu}$, the so called metric formalism \cite{felice}.\par
The variation of the determinant is always:
\begin{eqnarray}\label{vincent27}
\delta \sqrt{-g}=-\frac{1}{2}\sqrt{-g}g_{\mu\nu}\delta g^{\mu\nu}
\end{eqnarray}
and for the Ricci scalar, one has
\begin{eqnarray}
\delta R &=& \delta (g^{\mu\nu}R_{\mu\nu})\nonumber\\
&=&R_{\mu\nu}\delta g^{\mu\nu}+g^{\mu\nu}\delta R_{\mu\nu}\nonumber\\
&=&R_{\mu\nu}\delta g^{\mu\nu}+ g^{\mu\nu}\left(\nabla_{\sigma}\delta \Gamma^{\sigma}_{\nu\mu}-\nabla_{\nu}\delta\Gamma^{\sigma}_{\sigma\mu}\right)\,\,\label{vincent28}.
\end{eqnarray}
Now, since $\delta\Gamma^{\lambda}_{\mu\nu}$ is actually the difference of two connections, it should transform as a tensor. Therefore, one can write it as
\begin{eqnarray}\label{vincent29}
\delta\Gamma^{\lambda}_{\mu\nu}=\frac{1}{2}g^{\lambda\gamma}\left(\nabla_{\mu}\delta g_{\gamma\nu}+\nabla_{\nu}\delta g_{\gamma\mu}-\nabla_{\gamma}\delta g_{\mu\nu}\right)
\end{eqnarray}
and substituting in the equation (\ref{vincent28}) one finds
\begin{eqnarray}\label{vincent30}
\delta R= R_{\mu\nu}\delta g^{\mu\nu}+g_{\mu\nu}\Box\delta g^{\mu\nu}-\nabla_{\mu}\nabla_{\nu}\delta g^{\mu\nu}\,\,.
\end{eqnarray}
Hence, the variation in the action reads:
\begin{eqnarray}
\delta S&=&\frac{1}{2\kappa^2}\int\left[\delta f(R)\sqrt{-g}+f(R)\delta\sqrt{-g}\right]d^4x \nonumber\\
&=&\frac{1}{2\kappa^2}\int\left[ F(R)\delta R\sqrt{-g}-\frac{1}{2}\sqrt{-g}g_{\mu\nu}\delta g^{\mu\nu}f(R)\right]d^4x\nonumber\\
&=&\frac{1}{2\kappa^2}\int\sqrt{-g}\left[F(R)(R_{\mu\nu}\delta g^{\mu\nu}+g_{\mu\nu}\Box\delta g^{\mu\nu}-\nabla_{\mu}\nabla_{\nu}\delta g^{\mu\nu})-\frac{1}{2}g_{\mu\nu}\delta g^{\mu\nu}f(R)\right]d^4x\label{vincent31}
\end{eqnarray}
where $F(R)=\frac{\partial f(R)}{\partial R}$. Doing integration by parts on the second and third terms of (\ref{vincent31}), one gets
\begin{eqnarray}\label{vincent32}
\delta S= \frac{1}{2\kappa^2}\int\sqrt{-g}\delta g^{\mu\nu}\left\{F(R)R_{\mu\nu}-\frac{1}{2}g_{\mu\nu}f(R)+\left[g_{\mu\nu}\Box-\nabla_{\mu}\nabla_{\nu}\right]F(R)\right\}d^4x
\end{eqnarray}
Demanding that action remains invariant under variation of the metric, i.e $\delta S=0$, one obtains the field equations as
\begin{eqnarray}\label{vincent33}
F(R)R_{\mu\nu}-\frac{1}{2}f(R)g_{\mu\nu}-\nabla_{\mu}\nabla_{\nu}F(R)+g_{\mu\nu}\Box F(R)=\kappa^2T{\mu\nu}
\end{eqnarray}
where $T_{\mu\nu}$ is the energy momentum tensor of the matter fields defined by the variational derivative of $\mathcal{L}$ with respect to the metric $g^{\mu\nu}$:
\begin{eqnarray}\label{vincent34}
T_{\mu\nu}=-\frac{2}{\sqrt{-g}}\frac{\delta(\sqrt{-g}\mathcal{L})}{\delta g^{\mu\nu}}\,\,.
\end{eqnarray}
This satisfies the continuity equation
\begin{eqnarray}\label{vincent35}
\nabla^{\mu}T_{\mu\nu}=0\,\,,
\end{eqnarray}
The trace of (\ref{vincent33}) gives 
\begin{eqnarray}\label{vincent36}
3\Box F(R)+F(R)R-2f(R)=\kappa^2 T\,\,,
\end{eqnarray}
where $T=g^{\mu\nu}T_{\mu\nu}$ and $\Box F=\left(1/\sqrt{-g}\right)\partial_\mu\left(\sqrt{-g}g^{\mu\nu}\partial_\nu F\right)$. \par
Einstein gravity, without the cosmology constant, corresponds to $f(R)=R$ and $F(R)=1$, so that the term $\Box F(R)$ in (\ref{vincent36}) vanishes. In this case one has $R=-\kappa^2 T$ and hence the Ricci scalar is directly determined by the matter. In modified gravity the term $\Box F(R)$ does not vanish in (\ref{vincent36}). The time-time  and space-space components of the field equation (\ref{vincent33}) give respectively
\begin{eqnarray}
3FH^2-\frac{1}{2}\left(FR-f\right)+3H\dot{F}=\kappa^2\rho\,\,, \label{vincent37}\\ 
-2H\dot{F}-\ddot{F}-2F\dot{H}-3FH^2+\frac{1}{2}\left(FR-f\right)=\kappa^2 p\,\,. \label{vincent38}
\end{eqnarray}
Making use of (\ref{vincent25}) and combining (\ref{vincent37}) and (\ref{vincent38}) one obtains
\begin{eqnarray}\label{vincent39}
3(\omega+1)\left[FH^2+H\dot{F}-\frac{1}{6}\left(FR-f\right)\right]=H\dot{F}-2F\dot{H}-\ddot{F}\,\,.
\end{eqnarray}
The difference  with the modified gravity is that the usual matter  energy density is not the effective one. Then, even when there is no energy density for usual matter, there exists a kind of fluids called dark fluids \cite{felice}. This appears explicitly when we put (\ref{vincent37}) and (\ref{vincent38}) in the followings forms, neglecting the contributions of any other kind of usual matter,
\begin{eqnarray}
3H^2=\frac{1}{F}\left(\frac{1}{2}(FR-f)-3H\dot{F}\right) \label{vincent40}\\
-3H^2-2\dot{H}=\frac{1}{F}\left(\ddot{F}+2H\dot{F}-\frac{1}{2}(FR-f)\right) \label{vincent41}
\end{eqnarray}
Comparing Eqs. (\ref{vincent40}) and (\ref{vincent41}) with the standar Friedmann equations ($3H^2=\kappa^2\rho$ and $-3H^2-2\dot{H}=\kappa^2 p$), the both right sides of these equations may be identified  with the energy and pressure of a perfect fluid, in such a way that the barotropic  parameter of the EoS for this dark fluid is defined after simplification by
\begin{eqnarray}\label{vincent42}
\omega= \frac{p_{df}}{\rho_{df}}=\frac{-\frac{1}{2}(FR-f)+\ddot{F}+2H\dot{F}}{\frac{1}{2}(FR-f)-3H\dot{F}}\quad,
\end{eqnarray}
and the corresponding EoS may be written as follows:
\begin{eqnarray}\label{vincent43}
p_{df}=-\rho_{df}+\frac{1}{F\kappa^2}\left(-H\dot{F}+\ddot{F}\right)
\end{eqnarray}
where the subscript $"df"$ means "dark fluid". This inhomogeneous EoS (\ref{vincent43}) for this dark fluid takes the form of the kind of dark fluids studying in several works \cite{diego47, diego53, diego75, diego225, diego229}

One can put the function $f(R)$ in the Einstein-Hilbert part and a function $R^\alpha$ as $f(R)=R+\lambda R^\alpha$,  which includes the Starobinsky's model \cite{staro} as a specific case ($\alpha=2$),\, $\lambda$ being a constant.  Let us choose the scale factor as $a(t)=a_0t^r$, where  $r$ may be determined in terms of the barotropic parameter $\omega$ as usually done in the Einstein gravity. Then, using the curvature as $R=12H^2+6\dot{H}$,  (\ref{vincent39}) takes a new form as
\begin{eqnarray}
\frac{1}{6r(2r-1)t^2}\Bigg[-12r^2+42r^3-36r^4-18r^3(2r-1)\omega\nonumber\\
+\lambda\left(\frac{12r^2-6r}{t^2}\right)^\alpha t^2\left(-2\alpha+6\alpha^2-4\alpha^3+3r-5\alpha r+4\alpha^2r-6r^2+3\alpha r^2\right)\nonumber\\
+\lambda\omega\left(\frac{12r^2-6r}{t^2}\right)^\alpha t^2\left(3r-9\alpha r+6\alpha^2r-6r^2+3\alpha r^2\right)\Bigg]=0\,\,\,.\label{vincent44}
\end{eqnarray} 
Note that in the special case of $\alpha=1$ and setting $a_0=1$, one obtains the result of Einstein gravity, that is $r=\frac{2}{3(1+\omega)}$. Even for any $\alpha>1$, this results remains the same  with the ansatz $f= R+\lambda R^\alpha$. Remember that particle production  is directly connected with the curvature of the universe. Then, since the scale factor remains the same as is Einstein gravity, it is obvious that the curvature of the universe may remain the same and also as the effective mass. This means that particle creation  aspect is the same as in Einstein gravity when ordinary matter is taken into account. This result does not agree with one found by Pereira and collaborators \cite{ademir}. They used Palatini formalism in the context of f(R) gravity for analysing cosmological particle creation for a spatially flat universe and found that a conformal invariant metric does not forbid the creation of massless during the radiative era of the universe. Their result is that, in the context of modified gravity, the scale factor in radiative universe behaves as whose of de-Sitter universe in Einstein gravity which would lead to particle creation and consequence implies that Parker's result is valid only in the context of  general relativity. Note that in a radiative universe, the trace of the energy momentum vanishes and so, their equation (6) must lead to $f(R)\propto R^2$ which means that their equation (7) would never hold. From our result in this paper, it is clear that the scale factor of a radiative universe in the context of f(R) gravity remains the same as in Einstein gravity. Then, without any approximation, Parker's result must also hold in f(R) gravity. \par 
However, it is important to mention that models of the form $f(R)=R+\lambda R^\alpha$, that admit the existence of a viable matter dominated epoch prior to a late time acceleration requires  that the parameter $\alpha$ belongs to $\left] 0, 1\right[$,  \cite{amendola}. \par 
Let us consider that the main content of the universe is a dark fluid, so, neglecting any contribution of ordinary matter.  In the absence of ordinary matter fluid $(\rho=0)$, Eq (\ref{vincent37}) gives 
\begin{eqnarray}
3\left(1+\alpha\lambda R^{\alpha-1}\right)H^2=\frac{1}{2}(\alpha-1)\lambda R^\alpha - 3\alpha(\alpha-1)\lambda HR^{\alpha-2}\dot{R} \,\,\,.\label{vincent45}
\end{eqnarray}
   
The cosmic acceleration can be realized in the regime $F=1+\alpha\lambda R^{\alpha-1}>>1$ \cite{felice}. Under the approximation $F\approx \alpha\lambda R^{\alpha-1}$, by dividing Eq (\ref{vincent41}) by $3\alpha\lambda R^{\alpha-1}$, one obtains
\begin{eqnarray}
H^2\approx \frac{\alpha-1}{6\alpha}\left(R-6\alpha H\frac{\dot{R}}{R}\right)\label{vincent46}
\end{eqnarray}
Let us assume the power law expansion solutions for the scale factor, $a(t)=a_0t^r$, under the restriction that $r\neq 2$. The particular case, $r=2$, must be treated separately and will be presented later. With this power law expansion for the scale factor the Hubble parameter behaves as $H=r/t$ and using (\ref{vincent46}), we find
\begin{eqnarray}\label{vincent47}
r=\frac{(\alpha-1)(2\alpha-1)}{2-\alpha}
\end{eqnarray}
The acceleration requires that we have $r(r-1)>0$, which means that   
\begin{eqnarray}\label{vincent48}
\alpha\in I=]-\infty, \frac{1-\sqrt{3}}{2}[\,  \cup\, ]\frac{1}{2}, 1[\,\cup\,]\frac{1+\sqrt{3}}{2}, 2[\,\cup\,]2, \infty[\quad.
\end{eqnarray}
In terms of conformal time the scale factor behaves as
\begin{eqnarray}\label{vincent49}
a(\eta)\propto \eta^q, \quad q=\frac{r}{1-r}
\end{eqnarray}
As we need an expanding universe, one can distinguish  two different interval for the conformal time. Hence, for $0<r<1$ one has $0<\eta<\infty$ whereas for $r>1$ or $r<0$, one has $-\infty<\eta<0_-$.
We can now examine particle creation phenomenon in spatially flat spacetime  for massless scalar field minimally coupled with the gravitational field, $\xi=0$. Then, the equation (\ref{vincent5}) becomes 
\begin{eqnarray}\label{vincent50}
\chi_k^{\prime\prime}+\left[k^2-\frac{q(q-1)}{\eta^2}\right]\chi_k=0 \,\,\,.
\end{eqnarray}
This is essentially the same equation that governs the evolution of gravitational waves in an expanding universe \cite{gris}. This equation admits solutions in terms of hankel functions: 
\begin{eqnarray}\label{vincent51}
\chi_k(\eta)=\frac {\sqrt{\pi|\eta|}}{2}\left[A_kH^{(1)}_\nu(k|\eta|)+B_kH^{(2)}_\nu(k|\eta|)\right]\,\,, \quad \quad \nu=|1/2-q|\,\,\,,
\end{eqnarray}
where $A_k$ and $B_k$ are constants to be determined. Making use of the Wronskian relation
\begin{eqnarray}\label{vincent52}
zH^{(2)}_{\nu}(z)\partial_{z}H^{(1)}_{\nu}(z)-zH^{(1)}_{\nu}(z)\partial_{z}H^{(2)}_{\nu}(z)
= \frac{4i}{\pi} \quad ,
\end{eqnarray}
one obtains, imposing  orthonormalization of the modes, 
\begin{eqnarray}\label{vincent53}
\left|A_k\right|^2 - \left|B_{k}\right|^{2} = 1 \quad .
\end{eqnarray}
For fixing the initial vacuum state we use the Bunch-Davies sate \cite{article1, bunch} by the choice $B_k=0$ and $A_k=1$, then the solution (\ref{vincent51}) becomes
\begin{eqnarray}\label{vincent54}
\chi_k(\eta)=\frac {\sqrt{\pi|\eta|}}{2}H^{(1)}_\nu(k|\eta|)\,\,\,.
\end{eqnarray}
Note that the solution (\ref{vincent54}) reduces to the Minkowski one in which it does not occur particle creation phenomenon only for  the time interval $-\infty < \eta < 0_-$.\par
Now we can proceed to the calculation of $|\beta_k(\eta)|^2$ through the expression (\ref{vincent22}). Since we are dealing with a minimally coupled  massless scalar field,  the expression (\ref{vincent22}) becomes
\begin{eqnarray}\label{vincent55}
|\beta_k(\eta)|^2=\frac{1}{4}\frac{\left(\frac{q^2}{\eta^2}+k^2\right)|\chi_k |^2-\frac{q}{\eta}\left(\chi_k\chi^{\ast\,\prime}_k+\chi^{\prime}_{k}\chi^{\ast}_{k}\right)+|\chi^{\prime}_{k}|^2}{k^2|\bar{\chi}_k|^2}-\frac{1}{2}\quad.
\end{eqnarray}
Now, substituting (\ref{vincent54}) into (\ref{vincent55}) and making the change $\eta\rightarrow -\eta$, one we get
\begin{eqnarray}\label{vincent56}
|\beta_k(\eta)|^2=\frac{\pi}{8k}\eta^{-2q}\Bigg[\left[\frac{(q-1)^2}{\eta}+k^2\eta\right]H^{(1)}_{\nu}(k\eta)H^{(2)}_{\nu}(k\eta)\nonumber\\
-\frac{k}{2}(q-1)\left[H^{(1)}_{\nu}(k\eta)H^{(2)}_{\nu-1}(k\eta)+H^{(2)}_{\nu}(k\eta)H^{(1)}_{\nu-1}(k\eta)-H^{(1)}_{\nu}(k\eta)H^{(2)}_{\nu+1}(k\eta)-H^{(2)}_{\nu}(k\eta)H^{(1)}_{\nu+1}(k\eta)\right]\nonumber\\
+\frac{k^2\eta}{4}\left[H^{(1)}_{\nu-1}(k\eta)H^{(2)}_{\nu-1}(k\eta)-H^{(1)}_{\nu-1}(k\eta)H^{(2)}_{\nu+1}(k\eta)-H^{(1)}_{\nu+1}(k\eta)H^{(2)}_{\nu-1}(k\eta)+H^{(1)}_{\nu+1}(k\eta)H^{(2)}_{\nu+1}(k\eta)\right]\Bigg]
\end{eqnarray}

Note that at the early time, $\eta\rightarrow \infty$, $H^{(1)}_{\nu}(k\eta) \sim \sqrt{\frac{2}{k\pi \eta}}e^{i(k\eta-\frac{\pi}{4}-\frac{\nu\pi}{2})}$ and  using (\ref{vincent56}), we obtain $|\beta_k(\infty)|^2=0$. We see clearly through  (\ref{vincent56}) that particle production can be  known at any time and  also the initial vacuum condition is correctly reproduced at early time where there is no particle production. Then, as the conformal time grows, particle creation becomes important. Note that as the big rip time ($\eta \rightarrow 0_-$) is approached the rate of particle production diverges.  The evolution of the  particle production rate  for a $\alpha \in ]\frac{1}{2},1[$ is presented in Fig 1. We will see later that this behaviour of  particle production rate has an effective impact on the singularity.\par  
An important problem to be put out here is that, as the conformal time $\eta$ approaches $0_-$, the scale factor $a(\eta)$, the energy density $\rho_{df}$ and the pressure $p_{df}$ of the dark fluid, diverge: this corresponds to the big rip singularity.\par

Obviously, the question to be asked when there is appearance of singularity is : does this particle production, as quantum effects, can lead to the avoidance of such a singularity? To answer to this question it is important to evaluate quantum energy density and pressure,  and compare them  with the  classical ones of the dark fluid. The suitable expression for the quantum energy density is the renormalized one and this in perfectly known in the literature. For a minimally coupled scalar field, the renormalized energy density reads \cite{bunch}
 
\begin{eqnarray}\label{vincent57}
\left\langle T_{\mu\nu}\right\rangle_{ren}=\,\,-\frac{^{(1)}H_{\mu\nu}}{1152\pi^2}\ln{\left(  \frac{R}{6m^2}\right)} -\,\,\frac{^{(1)}H_{\mu\nu}}{1152\pi^2}\left[\psi\left(\frac{3}{2}
+\nu\right)
+\psi\left(\frac{3}{2}-\nu\right)\right]\nonumber\\
+\frac{1}{69120\pi^2}\Bigg(-168R_{\,;\mu\nu}+288\Box Rg_{\mu\nu}+24R_{\mu\sigma}R^{\sigma}_{\nu}\nonumber\\
-12R^{\alpha\beta}R_{\alpha\beta}g_{\mu\nu}-64RR_{\mu\nu}+63R^2g_{\mu\nu}\Bigg)
-\frac{Rg_{\mu\nu}}{192\pi^2C\eta^2}\,\,\quad.
\end{eqnarray}
Since we are leading with a massless scalar field, we need to take a null mass in (\ref{vincent57}). Note that in this massless limit, we have to introduce an arbitrary length (or inverse mass) $\mu$
\begin{equation}\label{vincent58}
   ^{(1)}H_{\mu\nu}\ln{\left(  \frac{R}{6m^2}\right)}=\,\, ^{(1)}H_{\mu\nu}\ln{\left(R\mu^2 \right)}-\,\,^{(1)}H_{\mu\nu}\ln{\left(6m^2\mu^2\right)}\quad.
\end{equation}
In a full gravitational dynamical theory, there would be a term $^{(1)}H_{\mu\nu}$ on the left-hand side of the gravitational field equations which arises from the presence on an $R^2$ term in the generalised gravitational action. Then, the final term of (\ref{vincent58}) is proportional to  $^{(1)}H_{\mu\nu}$ and so may be taken over the left-hand side of the field equations and absorbed in the renormalisation of the coupling constant of this term. The other term on the right side remains in $\left\langle T_{\mu\nu}\right\rangle_{\mbox{ren}}$ and is essential to the conservation of that quantity. Then, the energy momentum renormalised tensor of a massless minimally coupled scalar field is
\begin{eqnarray}\label{vincent59}
\left\langle T_{\mu\nu}\right\rangle_{ren}=-\frac{^{(1)}H_{\mu\nu}}{1152\pi^2}\ln{\left(R\mu^2\right)}
-\frac{^{(1)}H_{\mu\nu}}{1152\pi^2}\left[\psi\left(\frac{3}{2}
+\nu\right)
+\psi\left(\frac{3}{2}-\nu\right)\right]\nonumber\\
+\frac{1}{69120\pi^2}\Bigg(-168R_{\,;\mu\nu}+288\Box Rg_{\mu\nu}+24R_{\mu\sigma}R^{\sigma}_{\nu}\nonumber\\
-12R^{\alpha\beta}R_{\alpha\beta}g_{\mu\nu}-64RR_{\mu\nu}+63R^2g_{\mu\nu}\Bigg)
-\frac{Rg_{\mu\nu}}{192\pi^2C\eta^2}\,\,\quad.
\end{eqnarray}
With the scale factor  (\ref{vincent45}) we see from  (\ref{vincent59}) that in FRW the renormalised energy density behaves as (the same also occurs for the pressure)
\begin{eqnarray}\label{vincent60}
\rho_{ren} \propto \eta^{-4(q+1)}
\end{eqnarray}
For other hand, the energy density  of the dark fluid is
\begin{eqnarray}\label{vincent61}
\rho_{df} = \frac{1}{\kappa^2 F}\left[\frac{1}{2}(FR-f)-3H\dot{F}\right] \propto \eta^{-2\alpha(q+1)}\quad.
\end{eqnarray}
The ratio of the energy densities is 
\begin{eqnarray}\label{vincent62}
\frac{\rho_{ren}}{\rho_{df}} \propto  \eta^{2(\alpha-2)(q+1)}
\end{eqnarray}
Since we are leading with an expanding universe we have $\alpha\in I$ and two important cases can be distinguished: \par
$\bullet$ For  $\alpha \in J= ]-\infty, \frac{1-\sqrt{3}}{2}[\,\cup\,]\frac{1+\sqrt{3}}{2}, 2[\,\cup\,]2, \infty[$,  the product $(\alpha-2)(q+1)$ is positive and when the big rip time is approached , that is $\eta\rightarrow 0_-$,  the ratio of energy densities goes to zero. This means that the dark fluid is dominant and then the big rip cannot be avoided. \par

$\bullet$ For $\alpha\in\, ]\frac{1}{2}, 1[$, the product $(\alpha-2)(q+1)$ is negative and  the ratio of energy densities diverges. This indicates  that quantum effects may be dominant as the singularity is approached and the big rip may be avoided.  Note that this result agrees with that obtained in the \cite{article3} where it has been considered an expanding universe dominated by a dark energy fluid in  the Einstein gravity context. We also mention here that this is a case for which there is existence of viable model $f(R)=R+\lambda R^\alpha$, describing an early matter dominated and the late time acceleration universe.\par
However, it is important to mention that, the real aspect of the avoidance can be observed far before the singularity time. So, the question to be asked is: how long time before the big rip the quantum mechanism becomes important enough to begin dominating the matter content of the universe? It is a question that requires an analysis of many details of the evolution of the universe. But, we can answer this question in a semi-quantitative way. Let us put the general energy density due to quantum effects in the form
\begin{eqnarray}\label{vincent63}
\rho_{ren}=\rho_{ren 0}\eta^{-4(q+1)}\,\,,
\end{eqnarray}
where $\rho_{ren 0}$ is the energy density due to quantum effects today. On the other hand, the energy density of the dark fluid is given by $\rho_{df}=\rho_{df 0}\eta^{-2\alpha(q+1)}$. So, their ratio is
\begin{eqnarray}\label{vincent64}
\frac{\rho_{ren}}{\rho_{df}}= \frac{\rho_{ren 0}}{\rho_{df 0}} \left(\frac{\eta}{\eta_0}\right)^{2(\alpha-2)(q+1)}\,\,,
\end{eqnarray}
where $\eta_0$ characterizes the present conformal time. The scale factor is normalized so that it equals $1$ at present. Let us suppose that the present-time ratio between the two densities cannot exceed the ratio of the total radiation observed today with respect to the total density, i.e., $10^{-5}$. Hence the energy densities due to quantum effects and the dark fluid become comparable when
\begin{eqnarray}\label{vincent65}
\frac{\eta_e}{\eta_0} \sim 10^{\frac{5}{2(\alpha-2)(q+1)}}\,\,\,.
\end{eqnarray}
This may allow estimation of how much the scale factor has increased from today till the equality moment. Normalizing the present scale factor to unity and re-expressing it in terms of the cosmic time, we have
\begin{eqnarray}\label{vincent66}
a=\left(\frac{t_s-t}{t_s-t_0}\right)^{\frac{(\alpha-1)(2\alpha-1)}{(2-\alpha)}}\,\,,
\end{eqnarray}
where $t_0$ is the present cosmic time and $t_s$ the singularity time. Using the estimates made before, we find that the energy densities equal at
\begin{eqnarray}\label{vincent67}
t_s-t_e \sim 10^{-\frac{5}{2(2-\alpha)}}(t_s-t_0)\,\,,
\end{eqnarray}
where $t_e$ is the equality time. Hence, typically, the energy density due to quantum effects begins to dominate a fraction of the present age of the universe before the singularity. We conclude that it is sufficient to lead to big rip avoidance. So, quantum effects prevent the divergence which would occur about the energy density and pressure of the dark fluid as the singularity is approached and make them finite. This can be interpreted as a change of dynamical regime.

 As we told above, the case $\alpha=2$, the Starobinsky's model, must be analysed separately. Then, using directly the equation (\ref{vincent46}) without any ansatz about the explicit expression of the scale factor, one obtains 
\begin{eqnarray}\label{vincent68}
2H\ddot{H}+6\dot{H}H^2-\dot{H}^2=0  \,\,,
\end{eqnarray}
whose general solution cannot be found analytically. However, one can observe two particular solutions, $H=0$ and $H=const$, which can lead to a constant scale factor. With a constant Hubble parameter, the energy density and the pressure of the dark fluid are finite and then, there is no finite time future singularity. This result agrees with that obtained by Bamba and collaborators where they shown that the model
$R+\lambda R^2$ naturally removes finite time singularity \cite{bamba}.

\section{Cardy-Verlinde formula in f(R) context} 

This section is devoted to the application of the CV formula  to the dark fluid characterising the main contain of the universe. Note that the general approach to CV formula has already been done extensively for an ideal fluid in \cite{articlebrevik}. They derived CV formula-like which relates the entropy of the closed FRW universe to its energy, and Casimir energy, for a multicomponent coupled fluid where the generalized fluid obeys an inhomogeneous equation of state, both for Einstein and modified gravities. It has been explicitly shown as special result that when the equation of state is a linear one and the barotropic parameter $\omega=\frac{1}{3}$ (the radiative universe), the CV formula is recovered from the total entropy of the universe. Also, the recovery of the CV formula from Friedmann equations has been investigated  in our early work \cite{articleepl2} with Einstein gravity context, where the inhomogeneous equation of state contains a variable cosmological terms. The interesting case to be investigated here is the recovery of CV formula from the total entropy of the universe with an ideal fluid with inhomogeneous equation of state. As we know that the classical background goes toward a big rip, an analysis will be done around the singularity about CV formula terms.\par 
Looking at the equation (\ref{vincent43}), it clear that it is an inhomogeneous one. Then, as has been done in \cite{articlebrevik}, one can put it in general on the form 
\begin{eqnarray}\label{vincent69}
p_{df}=\omega(a)\rho_{df}+j(a)
\end{eqnarray}
and the conservation law for the energy,
\begin{eqnarray}\label{vincent70}
\dot{\rho}_{df}+nH(1+\omega)\rho_{df}=0\,\,\,,
\end{eqnarray}
where $n$ is the space dimension, becomes, using (\ref{vincent69})
\begin{eqnarray}\label{vincent71}
\frac{d\rho_{df}(a)}{da}+\frac{n(1+\omega(a))}{a}\rho_{df}(a)=-n\frac{j(a)}{a}\,\,\,.
\end{eqnarray} 
The general solution of (\ref{vincent71}) is 
\begin{eqnarray}\label{vincent72}
\rho_{df}(a)=e^{-G(a)}\left(Q-n\int e^{G(a)}\frac{j(a)}{a}\right)\,\,\,\,\,,
\quad G(a)=n\int^{a} \frac{1+\omega(a^{\ast})}{a^{\ast}}d a^{\ast}\,,
\end{eqnarray}
where $Q$ is an integration constant. Identifying (\ref{vincent43}) with (\ref{vincent71}), one gets
\begin{eqnarray}\label{vincent73}
\omega(a)=-1\,,\quad j(a)=\frac{1}{F\kappa^2}\left(-H\dot{F}+\ddot{F}\right)\,,
\end{eqnarray}
and consequently, $G(a)=0$.  Making use of  (\ref{vincent45}) and setting $Q=0$, one gets
\begin{eqnarray}\label{vincent74}
\rho_{df}(a)\propto a^{g(\alpha)}\,,\quad g(\alpha)= \frac{2\alpha(\alpha-2)}{(\alpha-1)(2\alpha-1)}
\end{eqnarray}
Hence, the total energy in the volume $V=a^n$ behaves as $E=\rho V \propto a^{n+g(\alpha)}$, which is also the behaviour of the extensive and sub-extensive energies. Assuming the conformal invariance, one can write the extensive and sub-extensive (the Casimir energy)  parts of the total energy as 
\begin{eqnarray}\label{vincent75}
E_E=\frac{b_1}{4\pi a^{-g(\alpha)-n}}S^{-\frac{g(\alpha)}{n}}\,,\quad E_C=\frac{b_2}{2\pi a^{-g(\alpha)-n}} S^{-\frac{g(\alpha)+2}{n}}\,,
\end{eqnarray}
 where $b_1$ and $b_2$, and $S$ the total entropy of the universe. Then the total entropy of the universe is 
 \begin{eqnarray}\label{vincent76}
 S=\left[ \frac{2\pi a^{-g(\alpha)-n}}{\sqrt{b_1b_2}}\sqrt{E_C(2E-E_C)}\right]^{-\frac{n}{g(\alpha)+1}}\,.
 \end{eqnarray}
Then, we see that the CV formula is recovered for $n+1=-g(\alpha)$, which means that the dimension of the spacetime  may be  $|g(\alpha)|$. Then, $g(\alpha)$ has to take only discrete values. Solving the equation $g(\alpha)+n+1=0$, one finds two value for the parameter $\alpha$,
\begin{eqnarray}\label{vincent77}
\alpha_1=1 \,,\quad  \quad \alpha_2=\frac{n+1}{2n+6}\,\,\,.
\end{eqnarray}
Then, it appears that CV formula may always be recovered from the total entropy of the universe in Einstein gravity. However, this occurs in f(R) gravity only for $\alpha=\alpha_2$.
On the other hand it is important to note that for either Einstein or f(R) gravity, if $n+1\neq -g(\alpha)$, CV formula cannot be recovered.\par
Let us now analyse what happens about CV formula terms around the big rip singularity. Note that the CV formula is reproduced for $g(\alpha)+n=-1$, then, the Casimir energy is inversely  proportional to the scale factor. We find then an interesting result that, in an expanding universe, the Casimir energy decreases and as the singularity (big rip) time is approached, the scale factor diverges and consequently the Casimir energy vanishes.  This result is in agreement with the  Brevik and collaborators's result where they shown that either with viscous or non-viscous fluid the Casimir effect fades away near the big rip \cite{brevikvisc}. On the other hand, when we put $\alpha_2$ on the form $\frac{n+1}{2(n+1)+4}$, it appears evidently that it is less that $1/2$ and then must belong to $J$. Hence the CV formula is reproduced only when the big rip cannot be avoided.

\section{Conclusion}
We studied particle creation aspect in f(R) gravity in which we considered the assumption $f(R)=R+\lambda R^n$. Note that particle creation is directly connected with the curvature of the universe. When the curvature vanishes, the modes are not affected by gravitational field and particle creation phenomenon cannot hold. This aspect is realized when we considered a massless scalar field in a radiative universe. Also in a static universe where the scalar remains constant at any time, the curvature vanished and particle creation cannot be expected. Note that these situation appears clearly when the field equation is put on the form of an harmonic oscillator equation. The frequency depends on the wave number $k$ and on the effective mass. Then, the gravitational field effect through the curvature is incorporated in the  effective mass. Consequently, when the curvature grows, the effective mass   does not vanish and the particle production can hold, at least when the mass is considered and/or  the coupling is not conformal. It appears that particle creation consequently is linked with the behaviour of the scale factor. We  found then that the scale factor does not change in the f(R) gravity with respect to his form in Einstein gravity and consequently the particle production aspect is the same as in Einstein gravity. We considered that the main content of the universe is a dark fluid and then the universe may evolve toward a finite time singularity, the big rip. We then analysed quantum effects near this singularity comparing the classical energy density and the pressure of the dark fluid with the renormalized energy density and the pressure due to quantum effects. Hence, we find that for $n \in ]\frac{1}{2}, 1[$ the ratio of the renormalized  energy density due to quantum effects to the classical energy density of the dark energy diverges as the big rip is approached. We conclude that  quantum effect are dominant near the singularity and the big rip may be avoided. For $n \in J$, the ratio goes do zero as the big rip is approached. This means that the dark fluid is the dominant component of the universe near the singularity and we conclude in this case that the big rip cannot be avoided. The Starobinsky's model has been analysed separately and reveals the absence of singularity. 
\par
However, due to the fact that power law expansion for the scale factor presents restriction on Starobinsky model, it would be interesting to use  an exponential solution and incorporating  conformal anomaly as quantum effects around the big rip. Also, in the case for which the singularity is not avoided, it would be interesting to allow the cosmic fluid to possess viscosity for analysing the possible avoidance of this singularity.  We propose to address these investigations in a future work.\par 
Another interesting point to which we devoted our attention in this work is the equivalence between the CV formula and the generalized total entropy of the universe, coming from the inhomogeneous equation of state that appears in the f(R) gravity. We found that the CV formula  can always be recovered in Einstein gravity while in f(R) gravity this equivalence holds only for $\alpha=\frac{n+1}{2n+6}$  corresponding to the situation in which the big rip cannot be avoided. We also analysed the behaviour of CV formula terms and found that as the universe expands, the Casimir energy decreases and fades away as the big rip is approached.

\vspace{1cm}

{\bf Acknowledgement:} The authors thank prof. S. D. Odintsov for useful criticisms and comments. 
M. J. S. Houndjo thanks  CNPq (Brazil) for partial financial support.
  A. V. Monwanou and J. B. Chabi Orou thank  IAEA/ICTP  STEP Program for financial support.
\newpage

\begin{figure}[htt!!!]
\centering
\includegraphics[height=12cm,width=15cm]{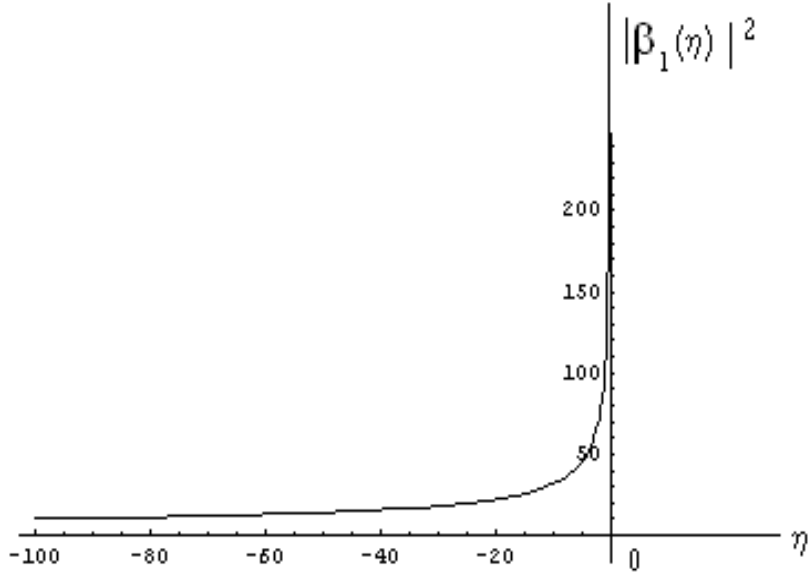}
\caption{\scriptsize{ The evolution of  particle creation rate as function of conformal time with the mode $k=1$ and $\alpha=2/3$ }}
\end{figure}

\end{document}